\documentclass[12pt]{iopart}

\usepackage{graphicx}

\newcommand {\be} {\begin{equation}}
\newcommand {\ee} {\end{equation}}
\newcommand {\bea} {\begin{eqnarray} }
\newcommand {\eea} {\nonumber \end{eqnarray}}

\newcommand {\ba} {\overline}
\newcommand {\lan} {\langle}
\newcommand {\ran} {\rangle}

\newcommand {\bc} {\begin{center}}
\newcommand {\ec} {\end{center}}
\newcommand {\bd}{\begin{displaymath}}
\newcommand {\ed}{\end{displaymath}}

\def \form#1 {eq. (\ref{#1}) }
\def \parziale#1#2  {{\partial {#1} \over \partial {#2}}}
\def \bi#1 {\typeout{#1} \item}

\begin{document}
 \title{Some considerations of finite dimensional spin glasses}

\author{Giorgio Parisi
}\address{Dipartimento di Fisica, Sezione INFN, SMC of INFM-CNR,\\
Universit\`a di Roma ``La Sapienza'',\\
Piazzale Aldo Moro 2,
I-00185 Rome (Italy)}
\ead{Giorgio.Parisi@INFN.Roma1.IT}
\begin{abstract}
In talk I will review the theoretical results that have been obtained for spin glasses, paying a particular attention to finite dimensional spin glasses.  I will concentrate my
attention on the formulation of the mean field approach and on its numerical and experimental verifications.  I will mainly considered equilibrium properties at zero magnetic
field, where the situation is clear and it should be not controversial.  I will present the various hypothesis at the basis of the theory and I will discuss their physical status.

\end{abstract}

\section{Introduction}

In this paper, after a very brief history of the subject, I will concentrate my attention on some of the main predictions of the replica approach \cite{MPV}-\cite{LH2006}, that
clearly distinguish it from other approaches (e.g. from the droplet model \cite{DROPLET} and from an intermediate possibility, the so called TNT scenario \cite{TNT}).  I will compare the
theoretical results with the data coming from simulations and briefly from experiments.

I will focus my attention on equilibrium results, not touching the very important region of off-equilibrium properties.  A main role has the study of the correlations,
both in the usual short range model and in the one dimensional model with long range interactions.  Finally I will discuss the status of the properties of ultrametricity in three
dimensional short models. Some conclusions are presented at the end.

\section{A brief history}

\subsection{Slow progresses}
Although the theoretical investigations of spin glasses have a long history, deep studies of spin glass theory started with the Edwards Anderson paper \cite{EA} and the Sherrington
Kirkpatrick papers \cite{SK}.  The SK papers made clear that there was something very strange in this {\em soluble} model \cite{DAT,TAP,Anderson}, whose {\em exact} solution was not
correct, as far it gave a negative entropy at low temperature.  This difficulty has been later solved \cite{MPV}, but the theoretical efforts have been quite large: they needed the introduction
of many new concepts.  The physical theory of complexity was born with this model.

The SK model is conceptually simple, it defines the mean field theory of spin glasses.
However the theory of spin glasses is difficult. Already in the mean field framework it took a long time to clarify the theory.
\begin{itemize}
\item  It took 5 years to solve a soluble model \cite{MPV}.

\item It took 5 years to find the full physical meaning of the physical solution after it was found \cite{MPV,PAR1}. 

\item It took other 10 years in order to found out the origin of some properties of the solution (e.g. stochastic stability \cite{Guerra,SS2,SS3}).

\item It took 25 years to prove that the free energy computation in mean field is correct \cite{V1,TALA}.

\item  It has not yet mathematically proved that all computations of the  observables done in mean field theory are correct, although there are no doubts on their correctness.
\end{itemize}

In finite dimensions we know the correlation functions in the Curie-Weiss approximation \cite{DKT}.  Already this computation is a {\em tour de force}.  The renormalization group
(i.e. the study of the non Gaussian fixed point) is still at the infancy.  A full computation at one loop of the corrections to the Curie-Weiss approximation is missing (logs are
presents in $D<6$).  Partial, but very instructive, computations have been done \cite{DKT} from which we have gained a great insight.  Ward identities have identified \cite{TD},
but we have not found a relevant non-linear sigma model that should give the dominant contribution at low temperature.  A simple argument has been presented that implies that the
lower critical dimension is $D=2.5$ \cite{INTERFACE}, in perfect agreement with the numerical data \cite{Boettcher}, but no field theory expansion around this dimension exists.
Moreover practically there is no real space renormalization group approach, that would be badly needed.

Why progress has been so slow?  We face is very far from being simple; in the course of the investigations we discovered new phenomena that
nobody were forecasting at the beginning of the investigations.  It turns out the order parameter in the mean field approximation is a probability distribution on an infinite
dimensional space, which sounds more or less likely a function of infinitely many variables \cite{V2,PAR1}: the free energy can be written as function of this probability
distribution.  In the actual solution of the SK model one can restrict oneself to consider only a very small subspace of these probability distributions, i.e. the stochastically
stable ultrametric distribution \footnote{A definition of stochastic stability and ultrametricity will be presented later.}.  Indeed the study of the SK model was the starting
point of analytic computations and theoretical understanding of complex systems in many different areas of science \cite{PAR1,Bakerian}.

\subsection{Some properties of mean field theory}
In mean field theory there are many equilibrium states (an infinite number of states in the infinite volume limit). These states are more or less equivalent and they can be distinguished one 
from the another by considering their mutual overlap
\be
q_{\alpha,\gamma}=N^{-1}\sum_{i=1,N}m(i)_{\alpha}m(i)_{\gamma},\ \ \ \ \ m(i)_{\alpha}\equiv \lan \sigma(i) \ran_{\alpha} \ ,
\ee
where $\lan \sigma_{i}\ran_{\alpha}$ denotes the expectation value restricted to the state $\alpha$.

The properties of these states (e.g. the set of the values of their overlaps) changes with the instance of the system. Analytically we can only compute their probability 
distribution, which is the order parameter of the problem as we have already seen.

For each system we define a function $P_{J}(q)$, $q$ being the overlap.  Its average is a function $P(q)$  defined as
\be
P(q)=\overline{P_{J}(q)}\ ,
\ee
where the overline denotes the average of the couplings $J$.

The function $P(q)$ is a not trivial also in finite dimensions.  Also the function
$P_{J}(q)$ is non trivial and it change from system to system.  All functions $P_{J}(q)$ have a delta peak at $q=q_{EA}\equiv q_{\alpha,\alpha}$ (the peak being obviously rounded
for a finite volume system).  The value of $q_{EA}$ in independent from the state and this and effect of their macroscopic indistinguishability.

Many phases are present for the same value of the parameters, therefore each point is a critical point (more precisely a multi critical point); sometimes (as it happens in the SK 
model) each
state is at a second order phase transition point and therefore long range correlations are expected to be present inside each state \cite{DK,CORRE}.  We have found here a
completely unexpected realization of SOCE (Self Organized Criticality at Equilibrium).

The educated reader will object that this picture is not possible for generic Hamiltonians.  This violates the Gibbs rule \cite{Gibbs} that states that multicritical points (where
$K+1$ phases are present) exist on a manifold of codimension $K$ in the parameter space.  Apparently this picture seems to be not consistent: it would be unstable under generic
perturbations unless something special happens.  However Guerra \cite{Guerra} was able to turn the argument around in a brilliant way: {\em something special does happen}.  The
whole picture is possible only if some conditions are satisfied, i.e. the so called stochastic stability identities found by Guerra.  They can be derived by assuming that the system is
stable against a generic perturbation.
\subsection{The two susceptibilities}
\begin{figure}
\begin{flushright}
	 \includegraphics[width=.53\textwidth]{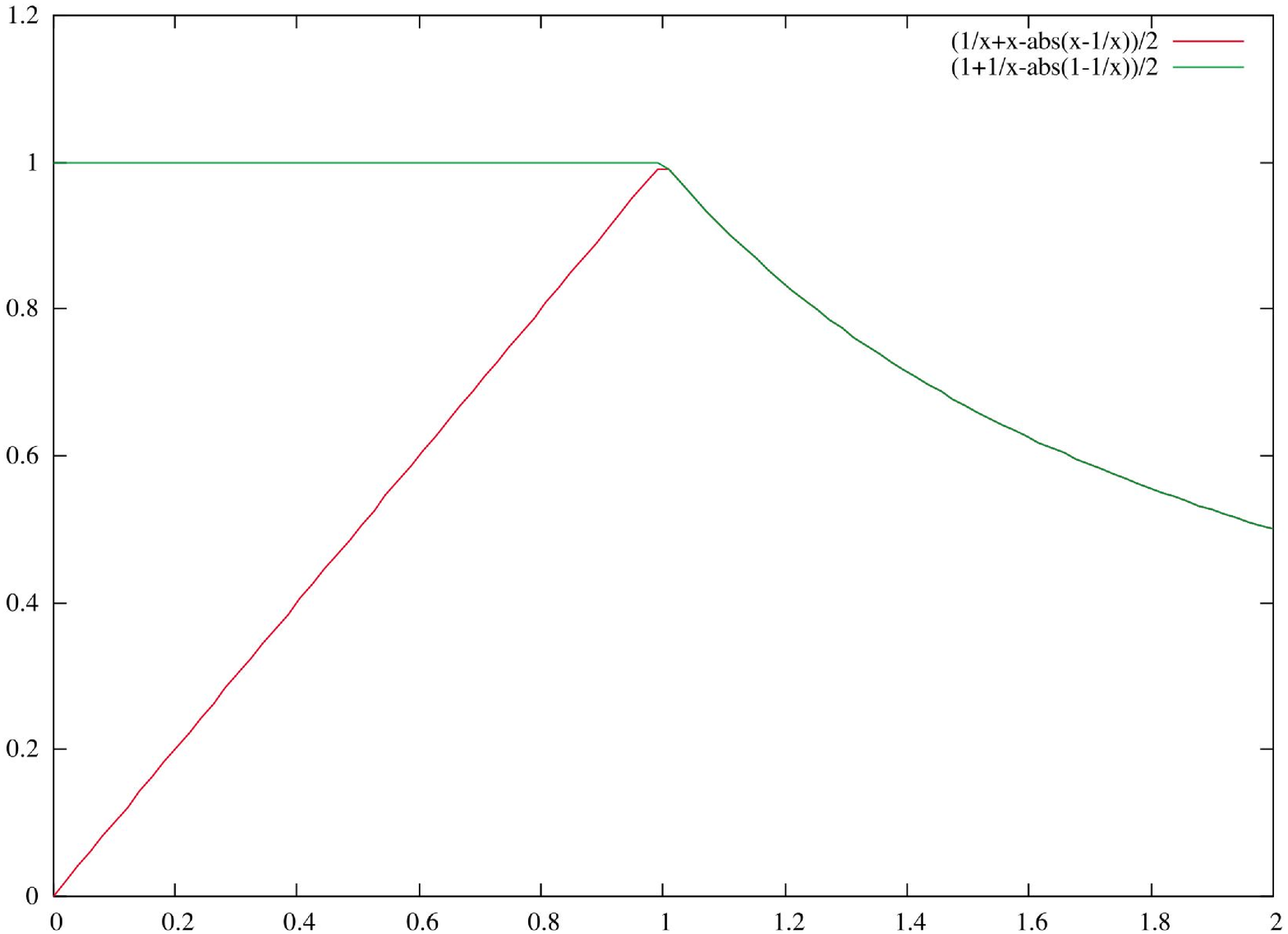}
	 \includegraphics[width=.45\textwidth]{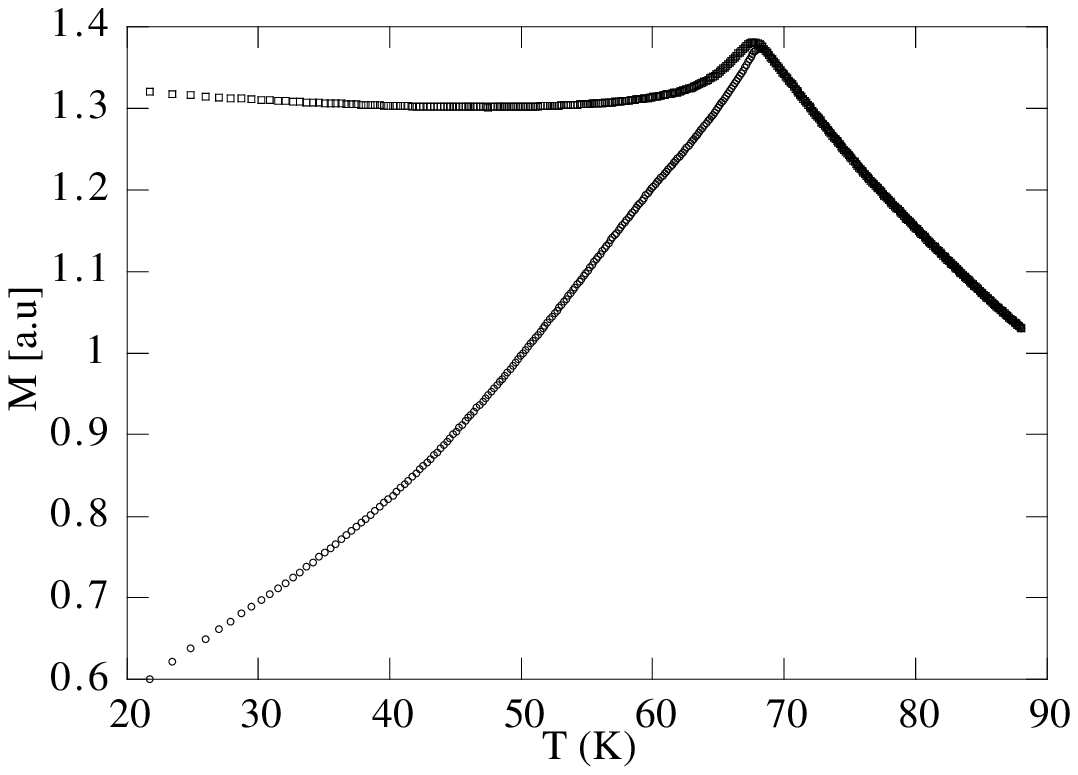}
\end{flushright}
\caption{At the left we obtain the results for the SK model, at the right we have experimental data on
	 metallic spin glasses \cite{EXP1}. The similarities among the two figures are striking.} 
	 \label{TWO}
\end{figure}

This picture has immediate experimental consequences.  We can define two physically relevant susceptibilities.
\begin{itemize}
	\item
The linear response susceptibility $\chi_{LR}$, i.e. the response within a state, that is observable when we change the magnetic field at fixed temperature and we do not wait too
much time \footnote{More precisely a time that is smaller than a function $f(h)$ that diverges when the magnetic field $h$ goes to infinity.}.  We have that
\be
\chi_{LR}=\beta(1-q_{EA})\ .
\ee
\item
The true equilibrium susceptibility $\chi_{eq}$, that is very near to $\chi_{FC} $, the field cooled 
susceptibility, where one cools the system in presence of a field:
\end{itemize}
\be
\chi_{eq}=\beta\left(1-\int |q|P(q)dq\right) \ .
\ee
The difference of the two susceptibilities is the hallmark of replica symmetry breaking. The experimental data are shown in fig. (\ref{TWO}).

\section{Present theoretical understanding}
\subsection{General considerations}
As we have already seen the actual solution of the SK model is a particular one, and in spite of its rather large complication, is the simplest one for a system with many 
equilibrium states. In principle one can consider more complex mean field solutions; sincerely I hope that these more complex solutions of the mean field equations are not 
necessary for spin glasses, but they could be useful in other contexts (e.g.  in non equilibrium systems  or in evolutionary systems with sexual reproduction \cite{PAR1}).

The  basic principles, on which the theory is based, are the following:
\begin{itemize}
	\item The function $P(q)$ is non-trivial and it changes from system to system.
	\item The system is stochastically stable, in presence of an infinitesimal magnetic field that breaks the spin reversal symmetry 
	\footnote{The spin reversal symmetry produces a twofold degeneracy of the equilibrium states, that disappears in an infinitesimal magnetic field). It would also possible to 
	eliminate the problem by considering $|q|$ at the place of $q$. See also the discussion in ref. \cite{CGGPV} on the sign of he product of the overlaps of three different 
	replicas.}.
	\item Overlap equivalence states that all other possible definitions of overlaps are equivalent to the original one. i.e. in the infinite volume limit  the new overlaps become  given 
	functions of the usual overlap $q$ \cite{SS3}.
	\item Ultrametricity, to be defined later.
\end{itemize}
This four principles are listed in order of relevance and generality. In principle it is possible that the last two could no be valid in sufficiently small dimensions and this would call for 
modifications of the field theory and (may be) for a more complex order parameter. 

It is clear that at sufficiently low dimensions (e.g. for dimensions less that 2.5) the order parameter must be zero and mean field does not apply anymore, however some remnants of
mean field theory may survive if the system is not too large or the observation time is not too long, as it happens for two dimensional superconductors where the Landau Ginsburg
theory applies very well at low temperatures on human scales, in spite of the absence of an order parameter at equilibrium.

\subsection{The probability distribution of the overlap}

\begin{figure}
 \includegraphics[width=0.45\textwidth]{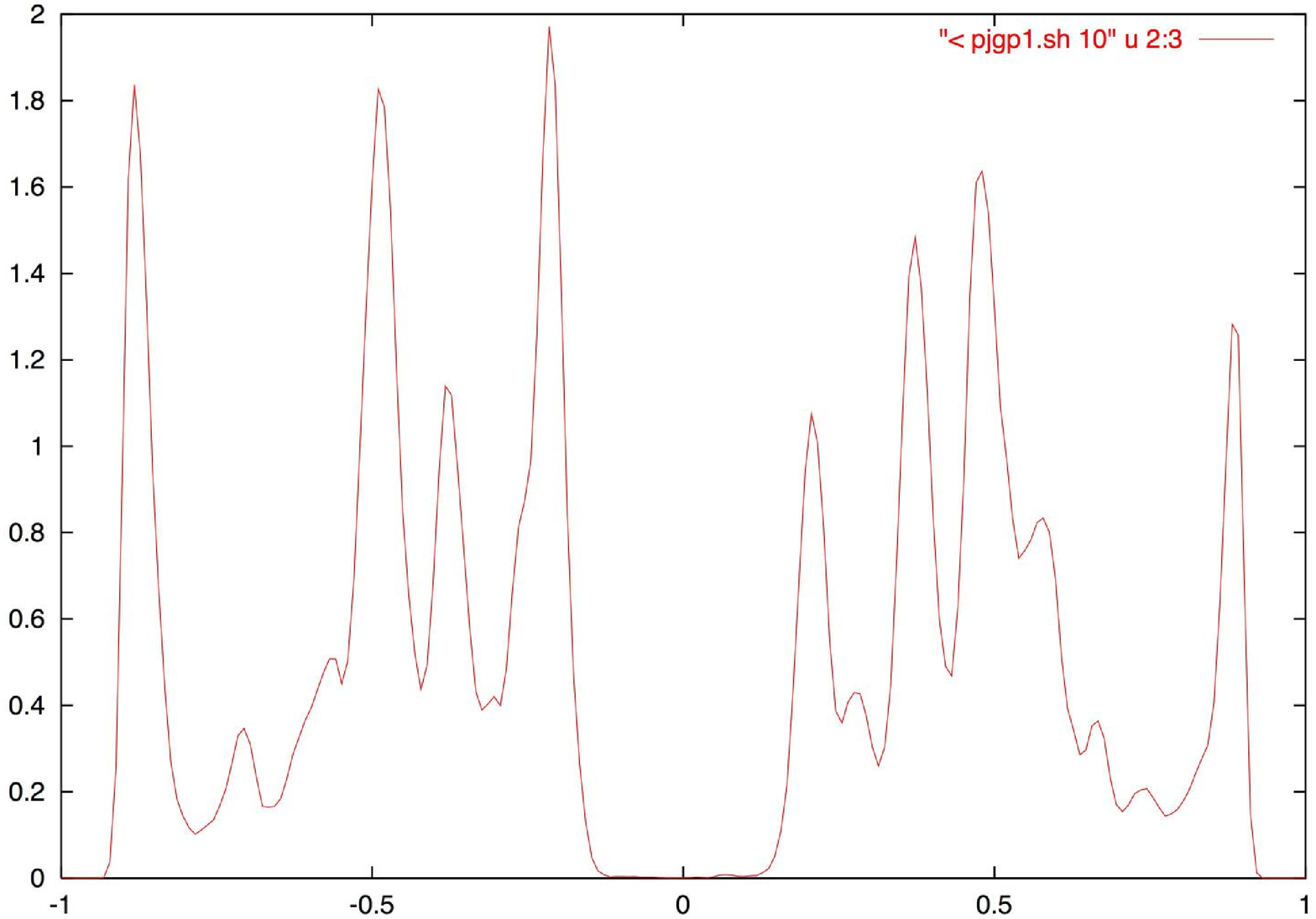}
  \includegraphics[width=0.45\textwidth]{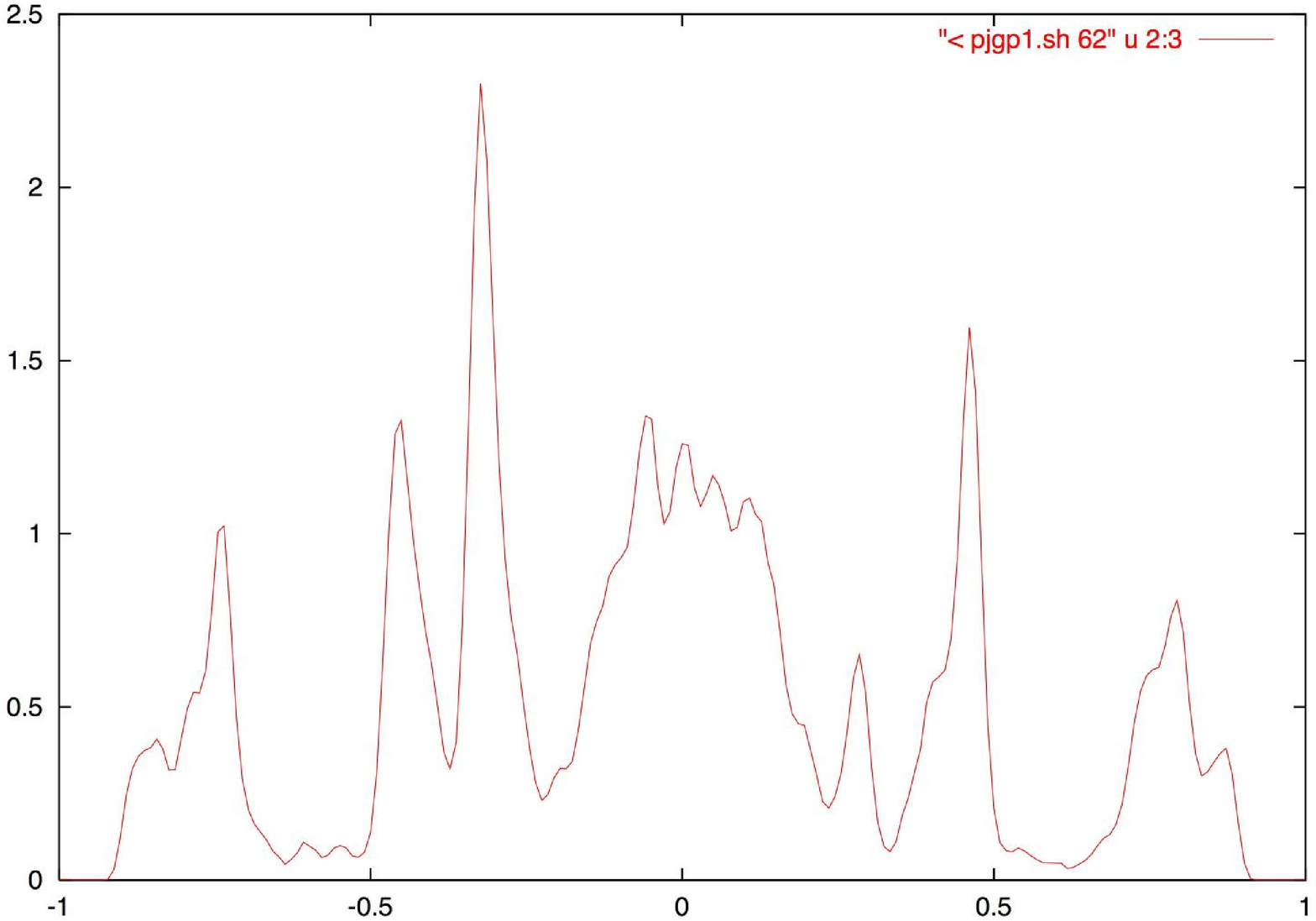}
\label{PQS}
\caption{The function $P_{J}(q)$ for two samples  
samples (i.e a different choice of $J$) for $D=3, \  L=16$ ($16^{3}$ spins)\cite{NUM1}.}
\end{figure}

\begin{figure}
  \includegraphics[width=.6\textwidth]{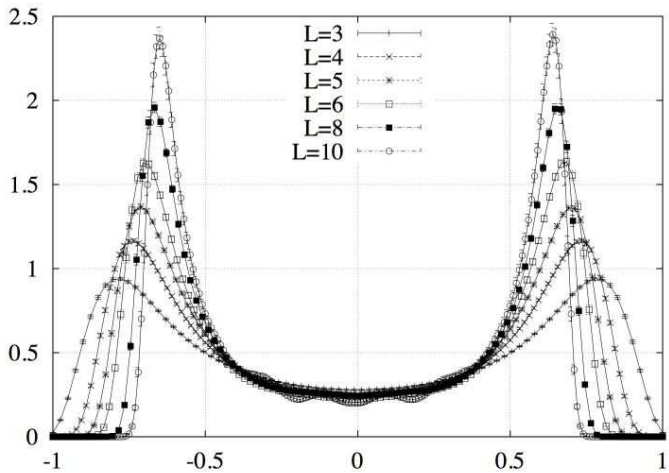}
\label{PQA}
\caption{The function $P(q)=\ba{P_{J}(q)}$ for $D=4$ after average over many samples (L=3\ldots 10) \cite{NUM2}.}
\end{figure}

There are no doubts from simulations, also in three dimensions (at least at zero magnetic field) that the function $P_{j}(q)$ is non trivial and  it fluctuates from system to system. In fig. 
(\ref{PQS}) we show some data for two three-dimensional samples with side $L=16$. In fig. 
(\ref{PQA}) we show the average function $P(q)$ in four dimensions.

This effect has been consistently seen in all the simulations for a quite large range of temperatures (practically up to $T=0$) and it persists also for  larger lattices (the largest 
lattices where systematic investigations have been done have side $L=24$ in dimensions 3. There are very strong indications that  this effect should persist in the infinite volume limit.
Also the critics of the replica approach accept this point, which is not controversial 
anymore.

\subsection{Stochastic stability.}

Stochastic stability is a very strong principle: it implies the existence of an infinite set of identities, maybe the most well known being:
\be
P(q_{1,2},q_{3,4})\equiv \overline{P_{J}(q_{1,2})P_{J}(q_{3,4})}=
\frac23  P(q_{1,2})P(q_{3,4}) +
	  \frac13  P(q_{1,2})\delta(q_{1,2}-q_{3,4}) \ .
\ee
The function $P(q_{1,2},q_{3,4})$ can be measured in simulations in a rather simple way by considering 4 identic replicas (or clones) of the the same system (i.e. all with the
same Hamiltonian).
A direct test of this last relation is not simple, because delta function are rounded in finite volume system.  Simpler relations are obtained by taking the moments.
For example let us consider four replicas of the system. The previous equation implies that
\begin{equation}
	       \ba{\lan q_{1,2}^{2} q_{3,4}^{2}\ran }    =
	  \frac23 \ba{\lan q_{1,2}^{2}\ran }^{2} +
	  \frac13  \ba{\lan q_{1,2}^{4})\ran } ,
\protect\label{MOM}
\end{equation}
which is satisfied numerically with very high accuracy \cite{CINQUE}.

In some sense stochastic stability is not an assumption.  If we consider the link overlap (to be defined later) at the place of the usual overlap, stochastic stability is a theorem
in equilibrium system that has been recently proved \cite{CONTUCCI}.  Of course the theorem is valid in the infinite volume limit and finite volume corrections have not been
evaluated, so that it is a welcome information to know that, at least as far the low moments are concerned as in equation eq.  (\ref{MOM}), the stochastic stabilities identities are
satisfied with high precision also for not too large systems.

Stochastic stability is a very strong requirement, whose implication have not completely spelled out.  We only notice here the very interesting fact that two non-interacting
stochastically stable system do not form a single stochastically stable system, so that stochastically stability implies in some sense the existence of a certain degree of
coherence of the system.  Moreover let me mention (although this paper is devoted to statics) that stochastic stability gives the link to relate the equilibrium properties to the
properties of systems that are slightly of equilibrium, recovering in this way the famous fluctuation dissipation relations \cite{Cuku1,Cuku2,FM,FMPP,OCIO}.  More detailed
consequences of stochastic stability are discussed in \cite{LOCO3,LOCO4}.
 
\subsection{Overlap equivalence}

Of course there could be many definition of overlaps \cite{PAR1}. You may take any quantity $O(i)$ and define an $O$-dependent overlap
as 
\be
q^{O}_{\alpha,\gamma}=N^{-1}\sum_{i}\lan O(i)\ran_{\alpha}\lan O(i)\ran_{\gamma}
\ee
A special case, that has been often investigated, is the link overlap, that on a $D$ dimensional lattice is defined as
\be
q^{L}_{\alpha_\gamma}=1/(DN)\sum_{i,k}\lan\sigma(i)\sigma(k)\ran_{\alpha}\lan\sigma(i)\sigma(k)\ran_{\gamma}
\ee
where the sums goes on all the $DN$ links $i$ $k$ of the cubic lattice (i.e. all the pairs of nearest neighbour points).

Overlap equivalence state that in the infinite volume limit 
\be
q^{O}_{\alpha,\gamma}=f^{O}(q_{\alpha,\gamma})\ , 
\ee
where the function $f^{O}$ may depend on the temperature. In the case of the SK model we have the simple relation:
\be
q^{L}_{\alpha_\gamma}=q_{\alpha,\gamma}^{2}\ .
\ee

This statement may be cast in a more  explicit way if we consider the overlap correlation $G(x)$ defined  in a system composed by two replicas of the same system (which for 
definiteness we call $\sigma$ and $\tau$):
\be
G(x)=\lan \sigma(x)\tau(x)\sigma(0)\tau(0) \ran \ .
\ee
We define also the constrained overlap correlation $G(x|q)$ where the previous average is done only on those pair of configurations that have overlap $q$.
It can be shown that overlap equivalence is equivalent to the statement that the correlation functions are clustering in the this constrained ensemble \cite{SS3} and therefore
\be
\lim_{x\to \infty}G(x|q)=q^{2}\ .
\ee
It is interesting to note that the correlation at distance 1, $G(1|q)$, is also called the link overlap $Q_{L}(q)$.
The previous relations implies that in the infinite volume limit
\be
\lan q^{O} \ran_{q} =f^{O}(q), \ \ \ \lan (q^{O})^{2} \ran_{q}=\lan q^{O} \ran_{q}^{2} \ .
\ee

In the same way in the ferromagnetic Ising model the violations of clustering and the presence of fluctuation in intensive quantities are eliminated by considering the 
constrained ensembles with positive (or negative) magnetization, therefore the sign of the magnetization is the order parameter. Here  the situation is similar, but the constraint 
is the value overlap. In the ensemble of two replicas at fixed overlap {\em intensive quantities do not fluctuate}. This principle partially closes the Pandora box opened by the 
fact that in the usual unconstrained ensemble intensive quantities do fluctuate \cite{MPV}.

This picture implies that the probability distribution of the window overlaps, i.e. overlaps in a region of side $R$ when the side $L$ goes to infinity first, is similar to that a 
system with size $L$, a crucial prediction that has 
been directly checked \cite{CINQUE,WINDOW}.

This scenario has been questioned and an alternative scenario, the TNT scenario, has been proposed \cite{TNT}.  According to this scenario the situation should be similar to the
Ising ferromagnetic with antiperiodic boundary conditions: there are many states (the interface may be anywhere) and there is a non trivial $P(q)$.  These states are locally
identical (apart from a spin flip) with the exclusion of a region whose relative volume goes to zero as $L^{-a}$ with $a=1$.  In the TNT scenario we would have that for $x<<L$,
$G(x|q)$ does not depend on $q$ for not too large $q$, i.e. for $1<<x<<L$
\be
G(x|q) =q_{ea}^{2} \ ,
\ee
while for  $x$ of $O(L)$, $G(x|q)$ is a function of $x/L$.

\begin{figure}
\includegraphics[width=.47\textwidth]{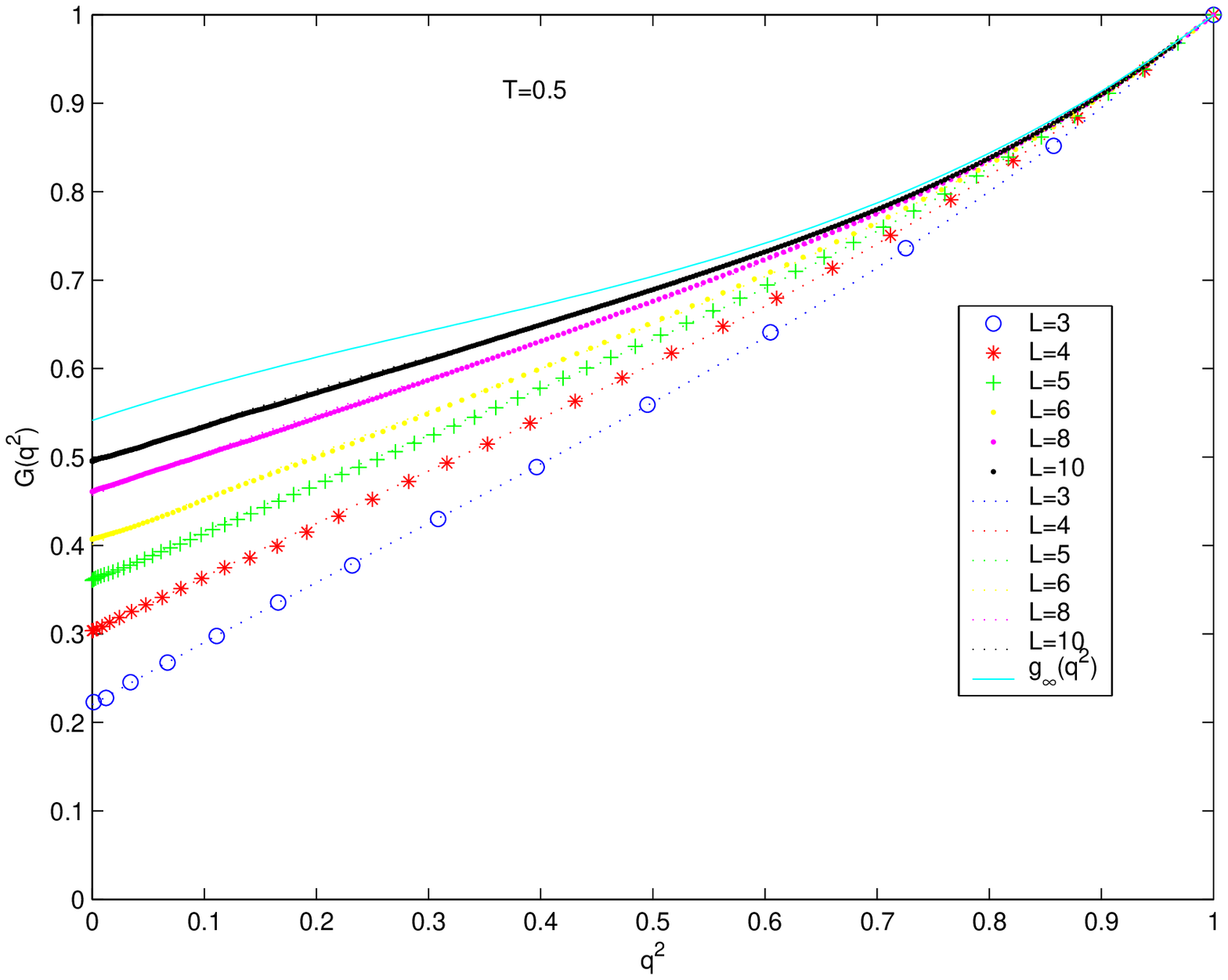} 
\includegraphics[width=.51\textwidth]{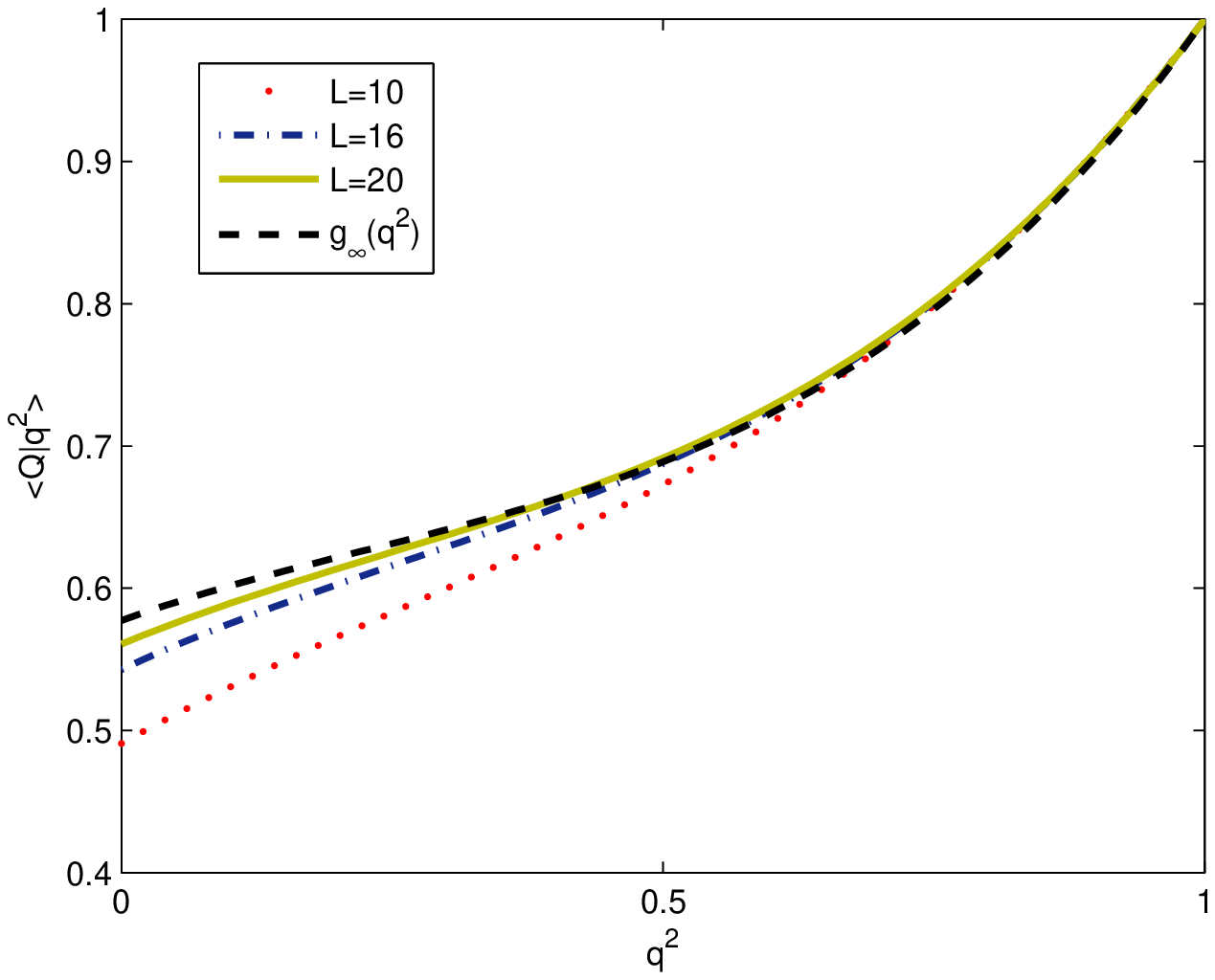}
\caption{
Plot of the curves $Q_{L}(q^2)$ ( for $T=0.5$ Gaussian model): Left panel $L=3,10$ and infinite volume extrapolation) \cite{CGGV}.
Right panel. Plot of the curves $Q_{L}(q^2)$ ( for $T=0.5$ bimodal model): $L=10,20$ and infinite volume extrapolation) \cite{CGGPV}.
}
\label{OVEEQUI}
\end{figure}

\begin{figure}
\includegraphics[width=.49\textwidth]{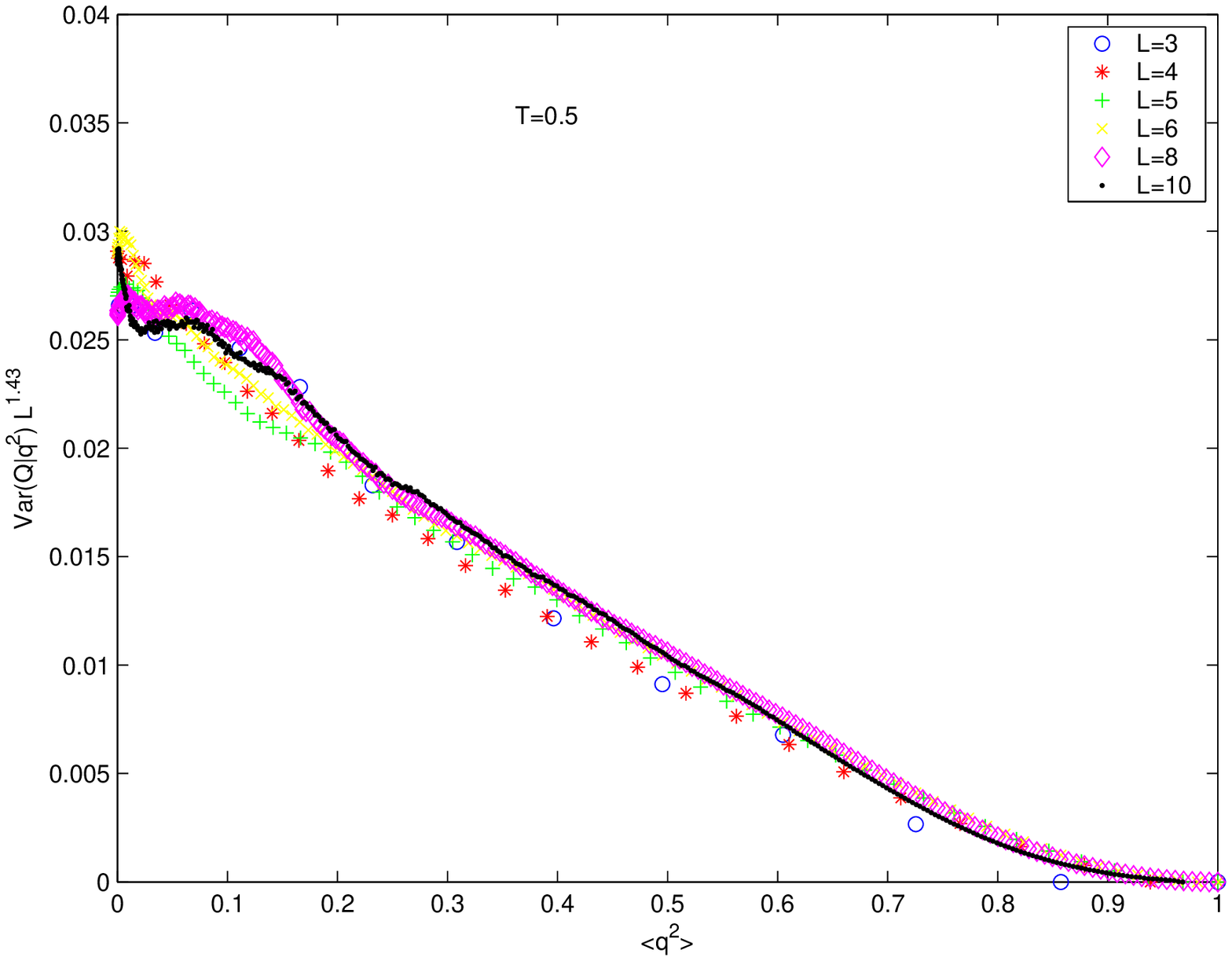}\includegraphics[width=.49\textwidth]{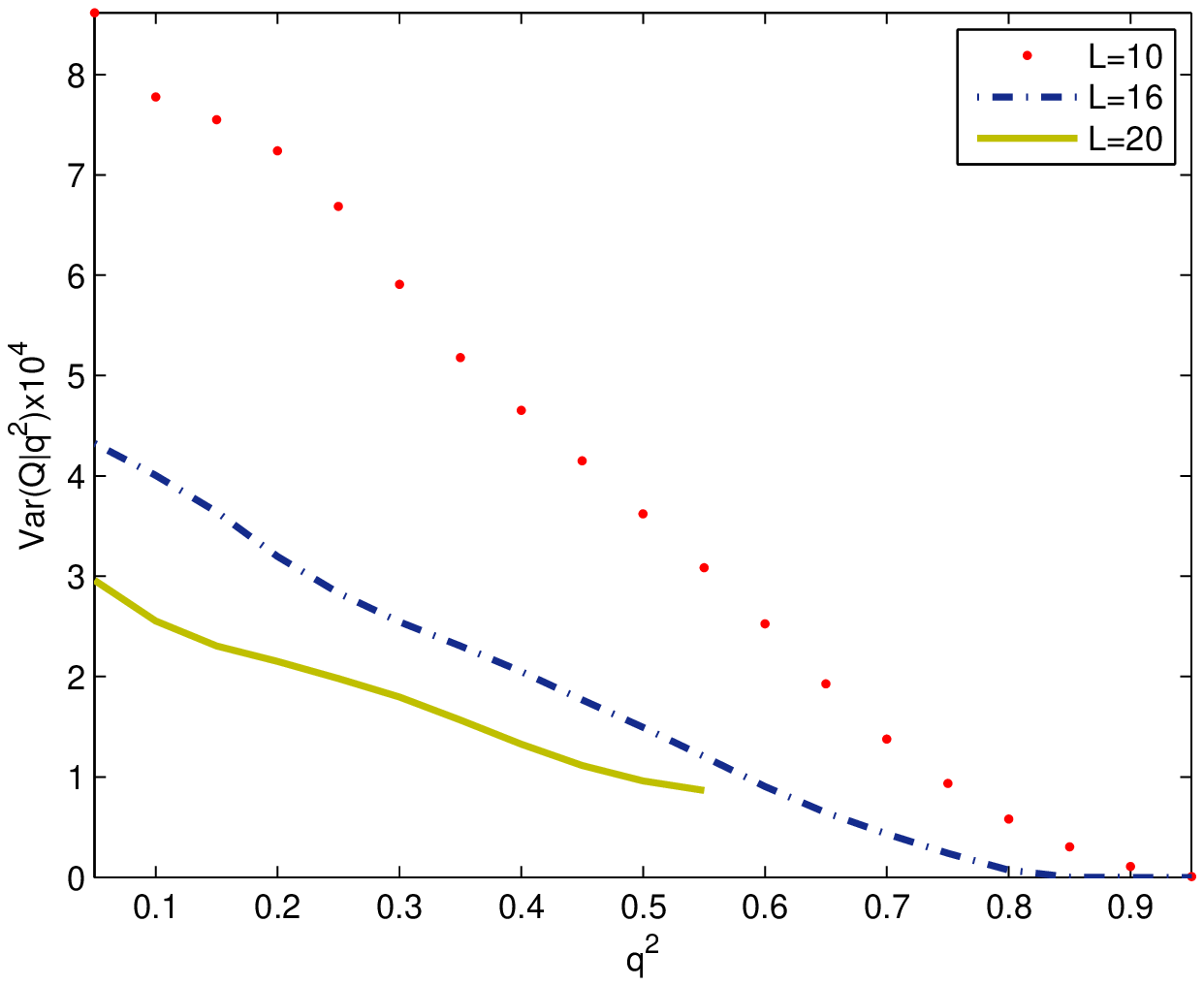}

\caption{Variance of $Q_{L}(q^2)$ (left panel divided by $L^{1.43}$ for the temperature $T=0.5$ \cite{CGGV}, right panel not divided by $L^{1.43}$ for the temperature $T=0.7$
\cite{CGGPV}), as a function of $q^2$, for different sizes $L$.}\label{VAR}
			   \end{figure}

The TNT scenario, in spite of the difficulties to put it in  a  stochastically stable form\footnote{The TNT scenario is also in variance with the known properties of the window 
overlaps \cite{CINQUE,WINDOW}.}, had some popularity a few years ago, when it was proposed for the first time.  It
would predict a delta function probability distribution for the link overlap.  However this conclusions were based on not too large lattices.  A more complete analysis already
showed that the arguments for the TNT scenario were related to transient finite volume effects \cite{MP}.  Indeed it now very clear, both from data for small lattices but at low 
temperature \cite{HD} and from data
on quite larger lattices (i.e. up to $L=20$) \cite{CGGV,CGGPV} that the TNT scenario is no correct and that for $q$ not very near to $q_{EA}$ we have
\be
\lan Q^{L}(q)\ran=A +B\, q^{2}\ ,
\ee
where $B$ has a definite limit when $L$ goes to infinity (see fig.  (\ref{OVEEQUI})).  Moreover the fluctuations of $Q^{L}(q)$, i.  e. its variance, go to zero quite fast with the
side (something like $L^{-1.4}$, see fig.  (\ref{VAR})).

\subsection{Dynamic correlations}

A particular interesting case are the correlations at  at $q=0$. According to the clustering properties they should go to zero at infinity, as a power with an exponent that can be 
estimated analytically:
\be
G(x|0) \propto x^{-\alpha}\ .
\ee

The interest on these correlations is that they can be computed in the dynamics \cite{DYNAMICS}: indeed if one consider a pairs of two large large systems, if $q=0$ at time zero,
$q$ remains zero at all times.  Therefore one can wait for a long time and compute the  equal time correlation function.  Obviously they will remain nearly equal to zero at distances
larger that the dynamically growing correlation length $\xi(t)$.  By definition in the limit $t \to \infty$ and fixed $x$ we obtain equilibrium correlations in some equilibrium state.

In the TNT scenario the correlation functions are the same in all equilibrium states in the infinite volume limit and for not to small $x$ they are given by $G(x) =q_{EA}^{2}$;
therefore we must have
\be
G(x,t) =q_{EA}^{2} f(x/\xi(t)) \ . \label{TNT}
\ee
On the contrary if overlap equivalence holds we should have
\be
G(x,t) ={x^{-\alpha}}  f(x/\xi(t))\ . \label{OE}
\ee
There is in the literature no indications whatsoever that eq. (\ref{TNT}) holds for the correlations; all numerical simulations (which have done in quite diverse 
situations) consistently support eq. (\ref{OE}).

However someone may object that in most cases the correlations functions are measured not at too large distances, so that a fit may be ambiguous.  A very interesting test of the
theoretical predictions can be done in one dimensional models with long range interactions.  If we use the diluted version \cite{LPRR} of Young long range model \cite{Y1}, one can
easily measure the
correlations functions at distances $10^{5}$ (linear lattices with $N=10^{6}$ sites can be simulated). The model is such that 
\be
\ba{J(i,)^{2}}\propto  |i-k|^{-\rho} \ .
\ee
More precisely in the diluted version $J_{i,k}=\pm 1$ with probability $|i-k|^{-\rho}$, otherwise it is zero \cite{LPRR}.

This model is a proxy of the $D$ dimensional short range model model, with the relation
$\rho=1+2/D$, in the sense that the two models have the same upper critical dimension, although they may have a different lower critical dimension.

\begin{figure}
\includegraphics[width=.49\textwidth]{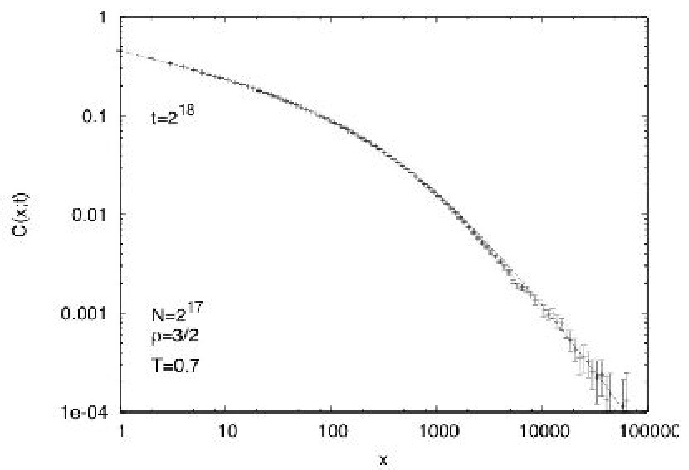}
	 \includegraphics[width=.49\textwidth]{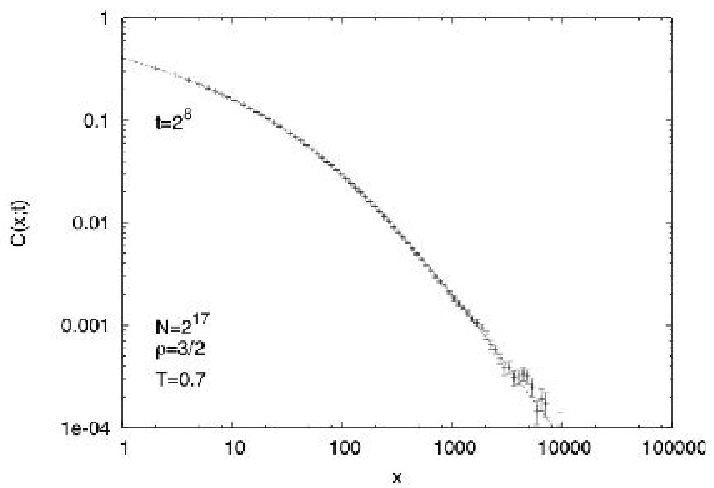}
\caption{The correlation functions on a one dimensional lattices with $N=2^{17}$ and $\rho=1.5$ in the low temperature phase ($T\approx .4 T_{c}$) at time $t=2^{18}$, left panel,
and at time $t=2^{8}$, right panel, as function of the distance, fitted with formula eq. (\ref{CROSS}).}
\label{FITS}
\end{figure}
The data  may be very well fitted for 4 decades in $x$ by
\be
G(x,t)=a x^{-\alpha} \left(1+(x/\xi(t))^{\delta(\rho-\alpha)}\right)^{-1/\delta} \ , \label{CROSS}
\ee
where we have imposed at very large distances $G(x,t) \propto x^{-\rho}$ (see for example fig. (\ref{FITS})).
It turns out that  $\alpha$ is roughly temperature independent below and not near to $T_{c}$. The correlation function (at least at some temperatures) seems to increase as stretched power.
i.e as $\xi (t) \propto B \exp(  A \log(t)^{1/2})$. 

The conclusion is quite clear: the dynamic correlation function behaviour in a way that is quite different from that of the droplet model and of TNT scenario and they are perfectly 
consistent with the clustering properties of the replica approach. Two different copies evolve locally toward two different incongruent ground states.

\section{Ultrametricity}

Ultrametricity states a very striking property
for a physical system: essentially it says that the equilibrium configurations of a large system
can be classified in a taxonomic (hierarchical) way (as animals in different taxa): configurations are grouped in
states, states are grouped in families, families are grouped in superfamilies. 

Ultrametricity implies that sampling three configurations  with
respect to their common Boltzmann-Gibbs probability, the distribution of the
distances among them is supported, in the limit of very large systems, only on equilateral and isosceles
triangles with no scalene triangles contribution.

We define  the three replicas overlap distribution ${\cal P}_{3}$.

\be
{\cal P}_3(q_{1,2},q_{2,3},q_{3,1}) =
\langle
%\sum_{\sigma,\tau,\gamma}
\delta(q_{1,2}-q(\sigma,\tau))\delta(q_{2,3}-q(\tau,\gamma))\delta(q_{3,1}-q(\gamma,\sigma))
\rangle \; ,
\ee
where $\sigma,\tau,\gamma$ denote three different equilibrium configurations. 

Ultrametricity implies that ${\cal P}_3$ is zero when the following inequality is violated
\be
q_{1,3}\ge \max(q_{1,2},q_{2,3})
\ee

{Stochastic stability implies \cite{PRzz} an exact formula for $\ba{{\cal P}_3}$:}
\bea
\ba{{\cal P}_3}(q_{1,2},q_{2,3},q_{3,1}) = 
\frac{X(q_{1,2}}{2} P(q_{12})X(q_{1,2})\delta(q_{1,2}-q_{2,3})\delta(q_{2,3}-q_{3,1})\nonumber
+\\ \nonumber
\frac12P(q_{1,2})P(q_{2,3})\theta(q_{1,2}-q_{2,3})\delta(q_{2,3}-q_{3,1})
\\
\frac12P(q_{2,3})P(q_{3,1})\theta(q_{2,3}-q_{3,1})\delta(q_{3,1}-q_{1,2})\\ \nonumber
+
\frac12P(q_{3,1})P(q_{1,2})\theta(q_{3,1}-q_{1,2})\delta(q_{1,2}-q_{2,3})  ,
\eea
where $X(q)=\int_{0}^{q}P(q)dq'$.

There are some indications that suggest that, for a stochastically stable system overlaps equivalence should imply ultrametricity \cite{ULTRANEC}, however no mathematical theorem
exists.

There were in the literature some direct test of ultrametricity in dimensions $4$ \cite{CMP} and some indications (coming from the dynamics \cite{FRT}) of the validity of
ultrametricity in D=3.
It is therefore interesting  to test directly if ultrametricity present in D=3.
\begin{figure}
\includegraphics[width=.59\textwidth]{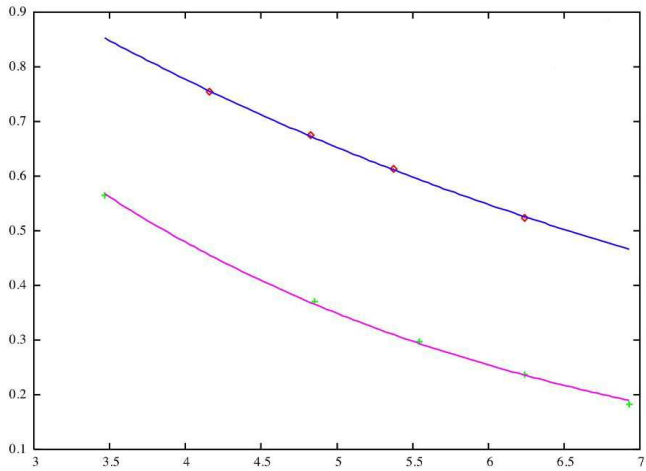}
\caption{Data extracted from \cite{CMP}: the average value of $K$ (including the point $K=1$, as function of the volume in the SK model (lower curve) and in the $3d$ EA model
(upper curve).  The curves are power fits to the data.}
\label{K}
\end{figure}

There have been some simulations in which the ultrametricity property was investigated \cite{HYD} by trying to identify the states and to define an ultrametricity index $K$ that takes
values in the interval $[0-1]$ such that the $K=0$ correspond to ultrametricity.  The simulations were done at low temperature in small lattices.

The expectation values of $K$ as function of the size of the system are shown in fig (\ref{K}) both for the SK model  and for the 3 dimensional Ising model. Both data are well 
fitted by a power of the volume 
\be
\lan K\ran \propto  N^{-\lambda}
\ee
with $\lambda=.31$ in the SK model and $\lambda=.17$ in $d=3$ (in other words  $K \propto L^{-.5}$). The data  suggest that $\lan K\ran $ goes to zero in the  infinite  
volume limit also in $D=3$ although for $L=8$, the maximum lattice explored the value of $K$ is not small. The reader should notice that these are {\em not }  the conclusions of the 
authors of the paper \cite{CMP}: they concentrated their attention on  the expectation of value of $K$ restricted to those configurations which have $K<1$ and with this 
restriction the behaviour is not so simple. Although the difference between the two quantities does not matters in the limit where 
$K$ goes to zero, the restriction  of considering only the data with  $K<1$ introduce a finite volume distortion and it may hide a simple power behavior.

As far as the extrapolation a $K=0$ starting from large values of $K$ could be dangerous, it is interesting to study directly the ultrametricity on a larger lattice, using also a
less sophisticated analysis that has a simple direct interpretation.

\begin{figure}
\includegraphics[width=.49\textwidth]{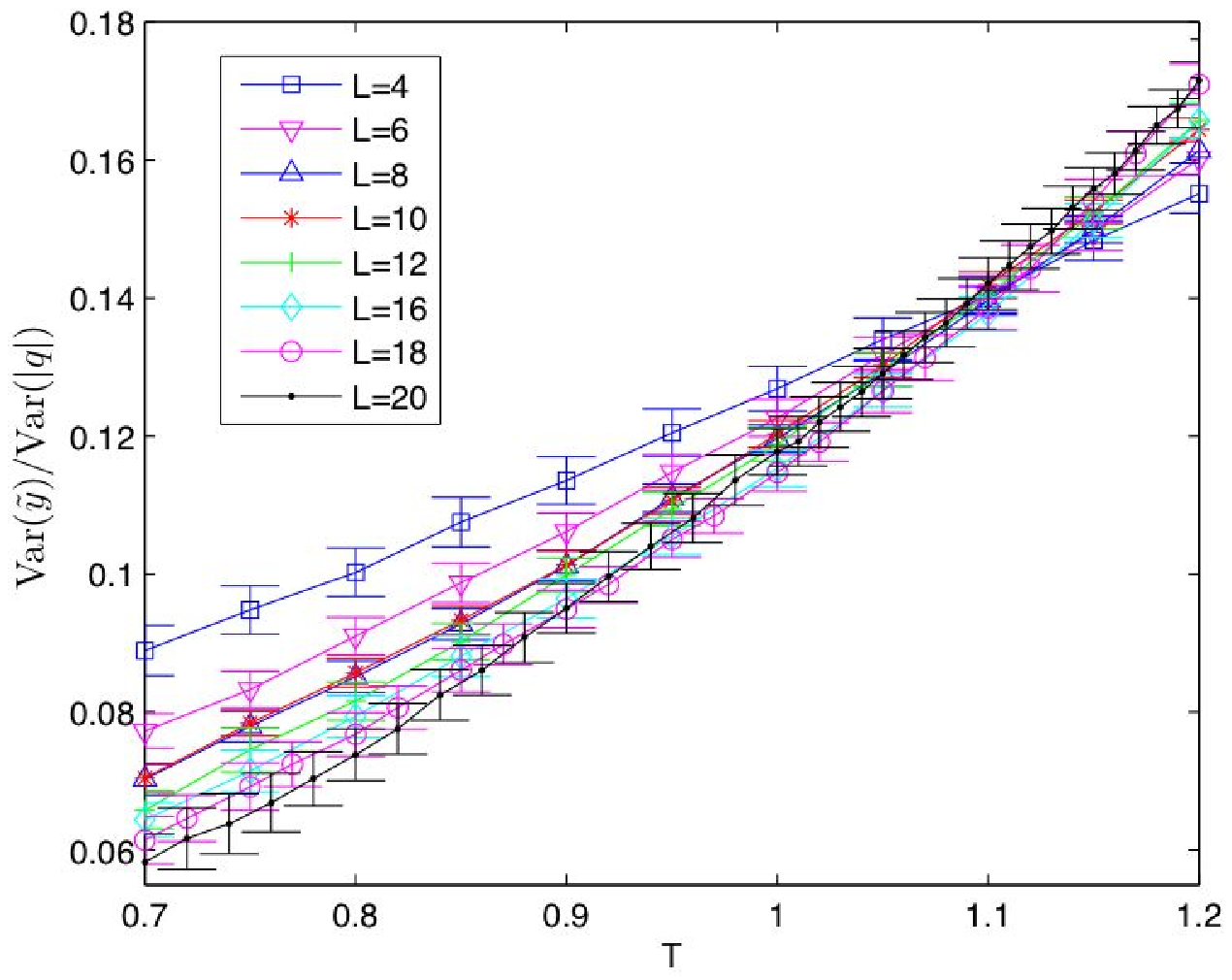}\includegraphics[width=.49\textwidth]{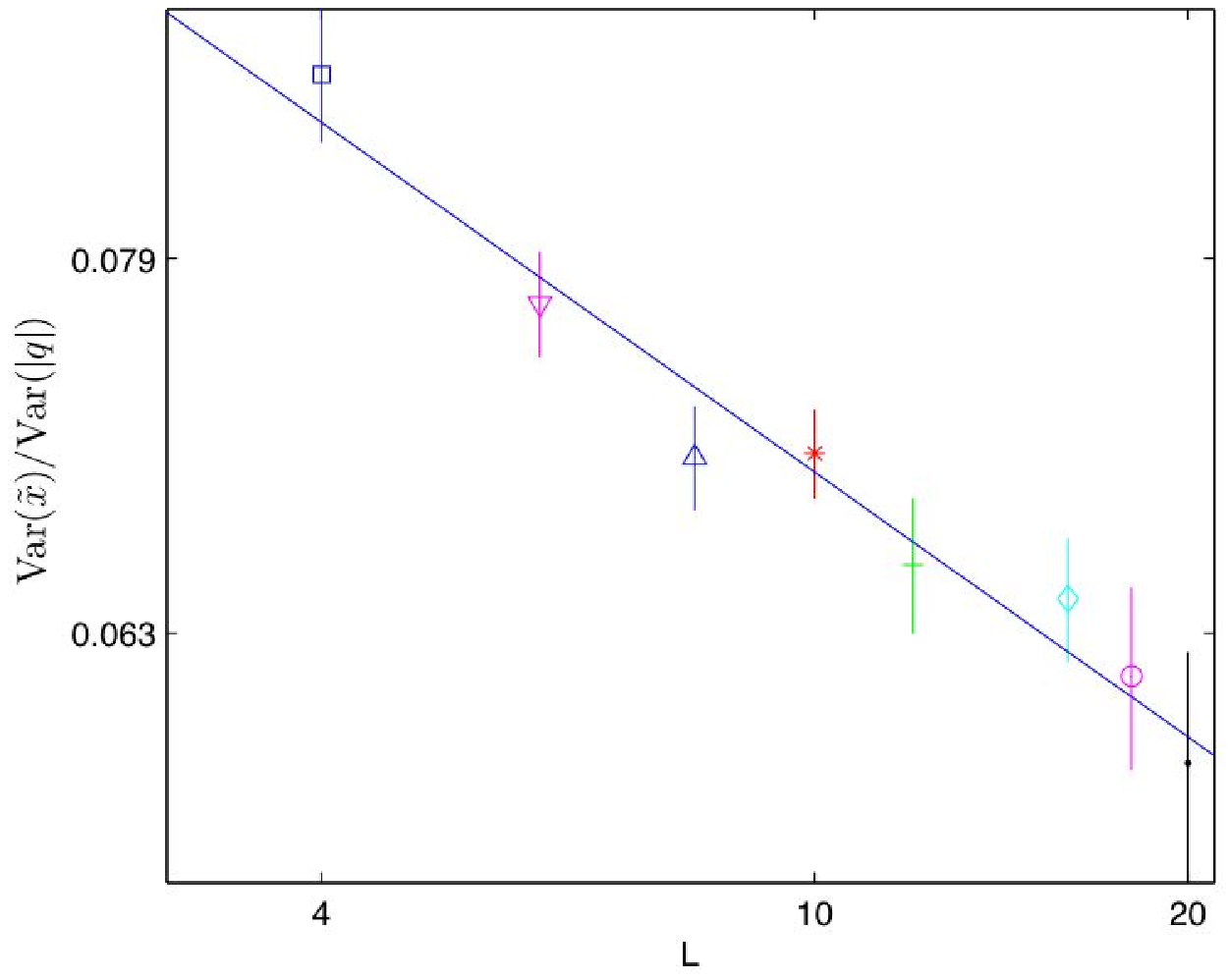}
\caption{The ultrametricity index $U$ as function of the temperature and of the volume (left panel); the curves crosses at the the critical  point. The ultrametricity index $U$ as 
function of the size of the lattice at $T=.7$ (right panel), where the line is a fit proportional to $L^{-.3}$. }
\label{ULTRA}
\end{figure}

This was done in \cite{CGGPV}. We considered  three independent configurations 1, 2, 3 such that 
\be
|q_{1,2}|<|q_{1,3}|<|q_{2,3}| \ ,
\ee
If we flip the configurations in such a way that $0<q_{1,3} < q_{2,3} $, it turns out that below the critical temperature $q_{1,2}$ is positive, or, if negative, very near to zero.
As far ultrametricity implies that $q_{1,2}=q_{1,3}$ a natural ultrametricity index is given by
\be
U=\frac{\lan (q_{1,2}-q_{1,3})^{2}}{\lan q^{2}\ran} \ .
\ee
We have measured $U$ for lattices going from $L=4$ to $L=20$ for temperatures up to $.6 T_{c}$, see fig. (\ref{ULTRA}).

At $T_{C}$ the probability distribution is not ultrametric but it nearly ultrametric: the value of $U$ is at the fixed point is much smaller that the value at the high temperature
Gaussian fixed point (i.e. around .8) Ultrametricity seems to be present below $T_{c}$, but it sets in slowly like $L^{-.3}$ .  In may be possible that the small power of $L$ is an
artifact of being too near to the critical temperature and that the exponent of the power may be slightly high at smaller temperatures, showing a more fast approach to
ultrametricity.

\section{Conclusions and questions to be answered}

Replica symmetry breaking picture is confirmed. Good evidence if found for overlap equivalence and alternative interpretative scheme like the droplet model or the $TNT$  scenario are 
no more viable.

Ultrametricity seems to present in the statics. However the measurements are difficult, there are strong finite volume corrections).  In the future, we must recheck the $D=4$ short
range model and study the long range 1D model for different values of $\rho$. Given the slow approach to ultrametricity in the static one may wonder what happens in the dynamics.
Can one use the measured violation of ultrametricity at finite volume in the statics to compute  the violations of dynamical ultrametricity in the dynamics?

A very important issue is the behaviour of spin glass in magnetic field. It is possible that in the same way that a random magnetic field destroys  the ferromagnetic order in D=2, 
a random (or constant) magnetic field may destroy spin glass order $D=3$.

In the last year some simulations papers appeared claiming that the de Almeida line, that should separate the replica exact from the replica broken regions is not present in D=3
\cite{Y1,Y2}.  I would be more prudent to reach  these conclusions:
\begin{itemize}
	\item For finite volume there is a cross-over region where configurations with negative magnetization have an important role also in presence of a small non zero magnetic 
	field.
	Any analysis done inside this region may be quite dangerous.
	\item There are analysis done on very large lattice with a different methodology that indicate that a transition is present in magnetic field and these analysis cannot be 
	easily dismissed \cite{DATS}.
	\item The previous analysis \cite{DATS} showed that in presence of a magnetic field the effects of replica symmetry breaking are strongly decreasing with decreasing the
	dimension, so is not clear if the analysis that miss the $dAT$ transition have enough sensitivity to detect it.
\end{itemize}

This problem should be carefully investigated and now we have all the tools to investigate the problem more deeply. It remains however a problem quite difficult to be studied, the 
window of magnetic fields (not too small, not to big) for which sensible results may be obtained may be small in simulations.

Generally speaking analytic progresses are also deeply needed: the development of a non-linear sigma model at low temperature and of a real space renormalization group results are
are very important steps that should be achieved.

We should be grateful to David for having given to us something interesting to work on not only in the last 30 year, but also for the next 30 years.

\end{document}